\documentclass[12pt]{article}
\pdfoutput=1
\usepackage{amsbsy}
\usepackage{pdfpages}
\usepackage{color}
\usepackage{amsfonts}
\usepackage{amssymb}
\usepackage{amsthm}
\usepackage{mathtools}
\usepackage{hyperref}
\usepackage{graphicx}
\usepackage{latexsym}
\usepackage[T1]{fontenc}
\usepackage{color}
\usepackage[titletoc]{appendix}
\usepackage{color}

\newcommand{\sect}[1]{\setcounter{equation}{0}\section{#1}}

\newcommand{\eq}{\begin{equation}}
\newcommand{\eqa}{\begin{eqnarray}}  
\newcommand{\en}{\end{equation}}
\newcommand{\ena}{\end{eqnarray}}
\newcommand{\enn}{\nonumber \end{equation}}

\def\Ahat{\widehat{A}}
\def\epsihat{{\widehat{\varepsilon}}}

\def\fhat{\widehat{f}}
\def\ghat{\widehat{g}}
\def\phihat{\widehat{\phi}}

\def\dehat{\widehat{\de}}

\def\sk{\vskip .4cm}
\def\noi{\noindent}
\def\om{\omega}
\def\al{\alpha}

\def\ga{\gamma}

\let \part\partial

\def\unquarto{{1 \over 4}}

\def\unmezzo{{1 \over 2}}
\def\epsi{\varepsilon}
\def\we{\wedge}

\def\de{\delta}

\def\part{\partial}

\def\sk{\vskip .4cm}

\def\noi{\noindent}

\def\X0{X^0}

\def\om{\omega}

\def\al{\alpha}
\def\ga{\gamma}

\def\unquarto{{1 \over 4}}
\def\unmezzo{{1 \over 2}}

\def\we{\wedge}

\def\de{\delta}

\def\A#1#2{ A^{#1}_{~~~#2} }

\def\square{{\,\lower0.9pt\vbox{\hrule \hbox{\vrule height 0.2 cm
\hskip 0.2 cm \vrule height 0.2 cm}\hrule}\,}}

\def\Lcal{{\cal L}}

\def\westar{\we_\star}

\def\omtilde{\tilde \om}
\def\Vtilde{\widetilde{V}}

\def\rtilde{\tilde r}
\def\epsitilde{\widetilde{\epsi}}

\def\psibar{\bar \psi}

\def\Om{\Omega}

\def\Ahat{{\hat A}}
\def\epsihat{{\hat \epsi}}
\def\dehat{{\hat \de}}
\def\Fhat{{\hat F}}
\def\fhat{{\hat f}}
\def\ghat{{\hat g}}
\def\phihat{{\hat \phi}}

\def\A{\cal A}
\def\of{\bar{\rm f}}

\def\dd{{\rm d}}

\def\FF{\mathcal F}
\def\st{\star}

\def\al{\alpha}

\def\ga{\gamma}

\def\sk{\vskip .4cm}
\def\noi{\noindent}
\def\om{\omega}
\def\al{\alpha}

\def\ga{\gamma}

\def\sk{\vskip .4cm}
\def\noi{\noindent}
\def\om{\omega}
\def\al{\alpha}

\def\ga{\gamma}

\let \part\partial

\def\la{\epsilon}

\def\unquarto{{1 \over 4}}

\def\unmezzo{{1 \over 2}}
\def\we{\wedge}

\def\de{\delta}

\def\part{\partial}

\def\sk{\vskip .4cm}

\def\noi{\noindent}

\def\X0{X^0}

\def\om{\omega}

\def\al{\alpha}
\def\ga{\gamma}

\def\unquarto{{1 \over 4}}
\def\unmezzo{{1 \over 2}}

\def\we{\wedge}

\def\de{\delta}

\def\square{{\,\lower0.9pt\vbox{\hrule \hbox{\vrule height 0.2 cm
\hskip 0.2 cm \vrule height 0.2 cm}\hrule}\,}}

\def\Lcal{{\cal L}}

\def\westar{\we_\star}

\def\omtilde{\tilde \om}
\def\Vtilde{\tilde V}

\def\rtilde{\tilde r}
\def\epsitilde{\tilde \epsi}

\def\psibar{\bar \psi}

\def\Om{\Omega}

\def\Ahat{{\hat A}}
\def\epsihat{{\hat \epsi}}
\def\dehat{{\hat \de}}
\def\Fhat{{\hat F}}
\def\fhat{{\hat f}}
\def\ghat{{\hat g}}
\def\phihat{{\hat \phi}}
\def\Psihat{{\hat \Psi}}

\def\*{\star}
\def\m*{\star^{op}}

\newcommand{\SWA}{\hat A}
\newcommand{\SWE}{\hat \epsilon}
\newcommand{\SWP}{\hat \phi}
\newcommand{\SWPA}{\hat \Psi}
\newcommand{\SWF}{\hat F}
\newcommand{\SWV}{\hat \delta}

\newcommand{\SWAp}{{\hat A}^{{\!\!\:\!\!\!\phantom{I}}^{{'}}}}
\newcommand{\SWEp}{{\hat \epsilon}^{\!\!\;'}}
\newcommand{\SWPp}{{\hat \phi}^{{\!\!\:\!\!\!\phantom{I}}^{{'}}}}
\newcommand{\SWVp}{{\hat \delta}^{{\!\;\!\!\!\!\phantom{I}}^{{'}}}}

\numberwithin{equation}{section}
\def\TT2{T}

\begin{document}

\begin{titlepage}
\vskip 2em
\begin{center}
{\Large \bf Noncommutative gauge and gravity theories\\[.2em] and geometric Seiberg--Witten map} \\[3em]

\vskip 0.5cm

{\bf
Paolo Aschieri and Leonardo Castellani }
\medskip

\vskip 0.5cm

{\sl Dipartimento di Scienze e Innovazione Tecnologica
\\Universit\`a del Piemonte Orientale, viale T. Michel 11, 15121 Alessandria, Italy\\ [.5em] INFN, Sezione di 
Torino, via P. Giuria 1, 10125 Torino, Italy\\ [.5em]
}
\end{center}
\phantom{.}\\[1em]

\begin{abstract}
\sk

We give a pedagogical account of noncommutative gauge and gravity theories, where
the exterior product between forms is deformed into a $\star$-product
via an abelian twist (e.g. the Groenewold--Moyal twist). The
Seiberg--Witten map between commutative and noncommutative gauge
theories is introduced. It allows to express the action of noncommutative Einstein gravity
coupled to spinor fields in terms of the usual commutative action with
commutative fields plus extra interaction terms dependent on the noncommutativity parameter.
\end{abstract}

\vskip 7cm
 \noi \hrule \vskip .2cm \noi {\small
paolo.aschieri@uniupo.it, leonardo.castellani@uniupo.it}

\end{titlepage}

\setcounter{page}{1}

\tableofcontents


\sect{Introduction}

Noncommutativity of phase space is a core feature of quantum mechanics.
Noncommutativity of spacetime rather than phase space has been
considered since the early days of quantum mechanics as a possible way to
reconcile gravity with quantum theory. 
Indeed the dynamical variable in Einstein general relativity is
spacetime itself with its metric structure, and noncommutativity of
spacetime coordinates could lead to a regularization in the perturbative treatment of
gravity as a quantum field theory. Noncommutativity of spacetime coordinates
is further supported by Gedanken experiments that aim at probing spacetime structures at very small distances. They show that due to gravitational backreaction one cannot test spacetime under the Planck scale. For example, in relativistic quantum mechanics the position of a particle can be detected with a precision of at most the order of its Compton wavelength $\lambda_C = \hbar/mc$. Probing spacetime at very short distances implies extremely heavy particles, that in turn produce high spacetime curvature. When $\lambda_C$ is of the order of the Planck length, the spacetime curvature radius has the same order of magnitude of the Compton wavelength of the probe particle, and the attempt to measure spacetime structure under the Planck scale fails.
Gedanken experiments of this type show that the description of
spacetime as a continuum of points (a smooth manifold) is an
assumption no more justified at the Planck scale. It is hence natural to
relax this assumption and conceive a more general noncommutative
spacetime, where uncertainty relations and discretization naturally
arise.  
Space and time are then described by a Noncommutative Geometry. In this way the impossibility of testing spacetime under the Planck
length, a dynamical feature due to gravitational backreaction, is
encoded at a deeper kinematical level.

In general spacetime discretization is expected in quantum gravity
theories, see the review \cite{Hossenfelder}.
For example in string theory the study of string scatterings leads to generalized
uncertainty principles where a minimal length emerges. In other
approaches, e.g. loop quantum gravity, minimal area and volumes are
predicted. 
\\

Spacetime noncommutativity also arises by considering an electron in a strong magnetic field $B$. In
this regime, due to the minimal coupling with the background gauge
field  $(A_x, A_y) = (0, Bx)$ associated to the flux $B$, the dynamics
takes place in the reduced phase space $q = y, p = Bx$. Thus the
electron’s coordinates become noncommutative: $[x, y] =
-\frac{i\hbar}{B}$ (cf. \cite{Jackiw:2001dj} for an extended discussion).
Hence quantum theory in the presence of a magnetic field $B$ leads to a noncommutative spacetime.
Similarly, low energy effective actions of open strings in the
presence of a background Neveu-Schwarz $B$-field can be described by
gauge theories on noncommutative spaces. 
The study of Yang-Mills (and Born-Infeld) theories on noncommutative spaces has proven very fruitful:
it allows to realize string theory T-duality symmetry within the low
energy physics of noncommutative (super) Yang-Mills theories \cite{Connes:1997cr}
It provides exact low energy D-brane effective actions, in a
given $\alpha'\to 0$ sector of string theory where closed strings
decouple, see ref. \cite{SW}. In that paper 
Seiberg and Witten provided an
explicit map (change of variables) between commutative and
noncommutative gauge theories.
\\

In this paper we study gravity on noncommutative spacetime as a noncommutative
gauge theory of the above type. The noncommutative geometry is first formulated in a
geometric (coordinate independent) language, useful for studying
diffeomorphism invariant theories. Then the specific gauge
theory describing noncommutative gravity in first order formalism
 (with independent vierbein and spin connection fields) is presented.
As we discuss in Section \ref{4.2Sec}, a generic feature of noncommutative gauge theories is that they are
well defined for $U(N)$ or $GL(N)$ gauge groups in the fundamental or
the adjoint representation but not for a generic representation, or
for a generic gauge group $G$ (like
e.g. $SU(N)$). This general feature implies that the Lorentz gauge
invariance of the first order gravity action becomes a $\star$-gauge
invariance that enlarges the classical $SO(1,3)$ group to $GL(2,C)$.

The enlargement of the gauge group corresponds to an increase in the
number of fundamental fields of the theory. This increase can be
mitigated by imposing charge conjugation constraints on the
noncommutative gauge action and can be fully avoided by the use of the
Seiberg-Witten map \cite{SW}, that relates the fields in the deformed
action (the ``quantum'' fields) to the classical fields, in such a way
that the ordinary gauge variations of the classical fields induce the
$\star$-gauge variations on the quantum fields. We thus obtain
  a gravity theory on noncommutative spacetime with the same
  degrees of freedom as classical gravity.

Among other approaches to noncommutative gravity we mention
gravity on fuzzy spaces \cite{Zoupanos}, emerging from matrix theory in
the presence of fuzzy extra dimensions \cite{Steinacker}, a metric approach 
\cite{NCG1}, \cite{NCG3} where the noncommutative
Levi-Civita connection is constructed and the braided gauge theory
approach of \cite{Ciric:2021rhi}.

Finally, a noncommutative hamiltonian formalism for twisted geometric theories
has been developed in \cite{LChamiltonian}, and applied to noncommutative vierbein gravity.
It allows an algorithmic construction of the canonical $\star$-gauge generators.

The plan of the paper is as follows. Section 2 deals with the origin
of twisted products, i.e. Weyl quantization and Groenewold--Moyal
product. Section 3 transfers noncommutativity of coordinates to
noncommutativity of functions and of exterior forms (via $\star$- wedge
products), and summarizes the basic results of the corresponding
noncommutative geometry. Section 4 illustrates the procedure in the
case of Yang-Mills theory, by deforming its classical action. In
Section 5 the $\star$-deformation of the gravity action is discussed
in detail reviewing \cite{AC1}. Section 6 contains a discussion on the Seiberg-Witten
map. In Section 7 it is shown how the Seiberg--Witten map allows to construct
noncommutative gauge theories with any gauge group. The
Seiberg--Witten map for the noncommutative gravity action is then
described in detail, providing a noncommutative action
with the same degrees of freedom as the commutative one \cite{AC2,
  ACD, Aext}. Expanding
this action in power series of the noncommutative deformation
parameter we obtain an action on commutative spacetime with
interaction terms dictated by noncommmutativity of spacetime.
As we further elaborate in the conclusions of Section 8, we
have thus constructed a modified gravity action that is expected to capture some quantum gravity aspects.

 \section{Noncommutative algebras, Weyl quantization and $\st$-products}
 
The easiest way to describe a noncommutative spacetime is via the
noncommutative algebra of its coordinates, i.e., giving a set of
generators and relations. We list three typical examples of commutation relations:
\begin{eqnarray}
\label{uuno}&[ x^\mu,  x^\nu]=i\theta^{\mu\nu}   ~~~~~~~~&{\mbox{\sl canonical}}\\[1em]
\label{ddue}&~~~~~[ x^\mu, x^\nu]=if^{\mu\nu}_{~~\sigma} x^\sigma   ~~~~~~~~&{\mbox{\sl Lie algebra}}
\\[1em]
\label{ttre}&~~~~~ x^\mu  x^\nu-q x^\nu  x^\mu=0   ~~~~~~~~&{\mbox{\sl quantum (hyper)plane}}
\end{eqnarray}
where $\theta^{\mu\nu}$ (a real antisymmetric matrix),
$f^{\mu\nu}_{~~\sigma}$ (real structure constants), $q$ (a complex
number, e.g. a phase) are the respective
  noncommutativity parameters. When the noncommutativity parameters are
turned off, the algebra becomes commutative and is the algebra of
polynomial functions on d-dimensional space $\mathbb{R}^d$. 
We can also impose further constraints, for example
periodicity of the coordinates describing the canonical noncommutative
spacetime  (\ref{uuno}). This leads to a
noncommutative torus rather than to a noncommutative
(hyper)plane. Similarly, constraining the coordinates of the  quantum (hyper)plane relations
(\ref{ttre}) we obtain a quantum (hyper)sphere. 
\sk
This algebraic description should then be complemented by a topological approach,
leading for example to the notion of continuous
functions. This is achieved by completing the algebra
generated by the noncommutative coordinates to a
$C^\star$-algebra. Typically $C^\st$-algebras arise as algebras of
operators on a Hilbert space. Connes noncommutative geometry \cite{connes} starts from
these notions and enriches the $C^\st$-algebra structure and its representation
on a Hilbert space, generalizing to the noncommutative case also the
notions of smooth functions and metric structure. 
\sk
A complementary approach to noncommutative space is given by the
$\star$-product, retaining the usual space of functions but deforming the pointwise product in a
noncommutative one. Historically the $\star$-product originated as a
noncommutative product for functions on phase space.
The quantization of phase space coordinates $q$, $p$ with Poisson
structure $\{q,p\}=1$ to operators  $\Phi(q)=\hat q$, $
\Phi(p)=\hat p$ on Hilbert space with $[\hat q,\hat p]=i\hbar$
is extended \`a la Weyl  to functions $f(p,q)\to \Phi(f)(\hat p, \hat
q)$. The operator product then induces a $\star$-product,
or Groenewold--Moyal  product, on functions
on phase space:
$$f\star g=\Phi^{-1}(\Phi(f)\:\!\Phi(g))~.$$
On polynomial functions Weyl quantization amounts to replace
$p, q$ with the operators $\hat p, \hat q$
and to symmetrize in $\hat p$ and
$\hat q$: $p^mq^n\mapsto \Phi(p^mq^n)=Sym(\hat p^m\hat q^n)$ where 
$Sym(\hat p^m\hat q^n)$ is the symmetrized polynomial in $\hat p, \hat
q$ normalized so that $Sym( p^mq^n)=p^mq^n$. It is defined by
$ (\hat p+\hat q)^{\ell}=\sum_{m+n=\ell} \frac{(m+n)!}{m!n!}Sym(\hat p^m\hat q^n).$
For example, $Sym(\hat p^m)=\hat p^m$, $Sym(\hat q^n)=\hat q^n$, $ Sym(\hat
p\hat q)=\frac{1}{2}(\hat p\hat q+\hat q\hat p)$,
$Sym(p^2q)=\frac{1}{3}(\hat p^2\hat q+\hat p \hat q\hat p+\hat q\hat p^2)$.
The corresponding $\star$-product explicitly reads $$(f\star g)(p,q)=
e^{\frac{i}{2}\hbar(\frac{\partial}{\partial q}
  \frac{\partial}{\partial p'}-\frac{\partial}{\partial p}
  \frac{\partial}{\partial q'})}f(p,q) g(p',q')\big|_{p=p',q=q'}~.$$ 
Since the operator product is associative so is the $\star$-product.

More in general (in formal deformation quantization) a $\star$-product on a manifold $M$ with Poisson structure $\{~\!,~\!\}$ 
is a noncommutative deformation of the usual pointwise product. It
sends two smooth functions $f,g$ to
a third one $f\star g$ and  is a differential operator on
both its arguments. It satisfies the
associative property $$f\st (g\st h)= (f\st
g)\st h~,$$ the normalization property $f\st 1=1\star f=f$ and
$$f\star g=fg+\frac{i}{2}{\hbar}\{f,g\}+\mathcal{O}(\hbar^2)$$ so that 
 in the semiclassical limit $\lim_{\hbar\to
  0}\frac{-i}{\hbar}(f\star g-g\star f)=\{f,g\}$, realizing the
correspondence principle betweeen quantum and classical mechanics. For
further reading on the topics of this section we refer to
\cite[Ch.~2, \S 3]{Takhtajan}, \cite[Ch. 6]{book}, \cite{revL}.

\section{$\star$-products from twists and noncommutative differential geometry}

The $\star$-product on phase space of the previous section generalizes to ${\mathbb{R}}^{d}$ with
coordinates $x^\mu$ as 
\eq\label{GMW}
( f\st g)(x)=
{\rm e}^{\frac{i}{2}\theta^{\mu\nu}{\partial\over\partial x^\mu}\otimes
{\partial\over\partial y^\nu}}f(x)h(y)\big|_{x=y}~.
\en
Here the antisymmetric matrix $\hbar \theta^{\mu\nu}$  has been for
short denoted  $\theta^{\mu\nu}$. Correspondingly, the classical limit $\hbar\to 0$
becomes $\theta\to 0$.

Notice that  if we set 
\[\FF^{-1}={\rm e}^{\frac{i}{2}\theta^{\mu\nu}{\partial\over\partial x^\mu}\otimes
{\partial\over\partial y^\nu}}
\]
then
\begin{equation}\label{muF-}
f\st g=
\mu(\FF^{-1}( f\otimes g))
\end{equation}
where $\mu$ is the usual product of functions $\mu(f\otimes g)=fg$.
The  element 
$\FF={\rm e}^{-\frac{i}{2}\theta^{\mu\nu}{\partial\over\partial x^\mu}\otimes
{\partial\over\partial y^\nu}}
$
 is an example of a Drinfeld twist. It is defined by the  exponential series in powers
 of the noncommutativity parameters $\theta^{\mu\nu}$,
\eq
\FF={\rm e}^{-\frac{i}{2}\theta^{\mu\nu}{\partial\over\partial x^\mu}\otimes
{\partial\over\partial y^\nu}}
=1\otimes 1 -{i\over 2}\theta^{\mu\nu}\partial_\mu\otimes \partial_\nu
-{1\over 8}  \theta^{\mu_1\nu_1}\theta^{\mu_2\nu_2}
\partial_{\mu_1}\partial_{\mu_2}\otimes \partial_{\nu_1} \partial_{\nu_2} + \ldots\nonumber
\en
It is easy to see that $x^\mu\star x^\nu-x^\nu\star
x^\mu=i\theta^{\mu\nu}$ thus  recovering  the
noncommutative algebra abstractly defined in (\ref{uuno}).

\sk
The method of constructing $\star$-products using Drinfeld twists
\cite{Drinfeld} (see e.g. \cite{GR2} for a quick introduction)
is not the
most general method (it does not apply to an arbitrary Poisson
manifold \cite{Weber}), however it is quite powerful,  and the class of
$\star$-products obtained is quite wide. For example choosing the
appropriate twist we can obtain 
the noncommutative relations (\ref{uuno}), (\ref{ddue}) and also (depending on the
explicit expression of the
structure constants)  some of the Lie
algebra type (\ref{ttre}).
It is also well adapted to a coordinate
free description of the $\st$-algebra of functions on a manifold $M$
and to its differential geometry.
\\

Let $M$ be a smooth manifold. A twist  is an invertible element $\FF\in U\Xi\otimes U\Xi$
where $U\Xi$ is the universal enveloping algebra of vector fields, (i.e. it
is the algebra generated by vector fields on $M$ and where the element
$XY-YX$ is identified with the vector field $[X,Y]$). The element $\FF$
must satisfy some further conditions that we do not write here, but
that hold true if we consider abelian twists, i.e., twists of the form
\begin{equation}\label{Mtwist}
{\mathcal F^{}}={\rm e}^{-\frac{i}{2}\theta^{IJ}
{X_I}\otimes
{X_J}}
=1\otimes 1 -{i\over 2}\theta^{IJ}X_I\otimes X_J
-{1\over 8}  \theta^{I_1J_1}\theta^{I_2J_2}
X_{I_1}X_{I_2}\otimes X_{J_1} X_{J_2} + \ldots
\end{equation}
where the vector fields $X_I$ ($I=1,...s$ with $s$ not necessarily equal
to $d={\rm{dim}} \,M$) are mutually commuting $[X_I,X_J]=0$ (hence
the name abelian twist).

It is convenient to introduce the following notation
\begin{equation}\begin{split}
\mathcal{F}^{-1}
&=1\otimes 1 +{i\over 2}\theta^{IJ}X_I\otimes X_J
-{1\over 8}  \theta^{I_1J_1}\theta^{I_2J_2}
X_{I_1}X_{I_2}\otimes X_{J_1}  X_{J_2} +\ldots\nonumber\\
&=\of^\alpha\otimes\of_\alpha\label{2.3eq}
\end{split}\end{equation}
where a sum over the multi-index $\alpha$ is understood.
\\

With a twist $\FF$ we deform the whole differential geometry of
$M$. Let $\A$ be the algebra of smooth functions on the manifold $M$. We deform $\A$ to a noncommutative algebra
$\A_\st$ by defining the new product of functions
$$f\st g=\of^\al(f)\,\of_\al(g)\,.$$
We see that this formula is a generalization of the Gronewold--Moyal star
product on ${\mathbb{R}}^{d}$ defined in (\ref{GMW}) or \eqref{muF-}.
Since the vector fields $X_I$ are mutually commuting, this
$\star$-product is associative.
Note that only the algebra structure of $\A$ is changed to $\A_\st$
while, as vector spaces, $\A$ and $\A_\st$ are the same.
We similarly consider the algebra of exterior forms $\Omega^\bullet$ with the wedge
product $\wedge$, and deform it in the noncommutative
exterior algebra $\Omega_\st^\bullet$ that is characterized by the graded
noncommutative exterior product $\wedge_\st$  given by
$$\tau\wedge_\st \tau'=\of^\al(\tau)\wedge \of_\al(\tau')~,$$
where $\tau$ and $ \tau'$ are arbitrary exterior forms, and each vector field  $X_{I_1}, X_{I_2}, X_{J_1},
X_{J_2}\ldots$ in (\ref{2.3eq}) acts on forms via the Lie derivative.
Only the product is deformed and hence $\Omega_\st^\bullet=\Omega^\bullet$ as
(graded) vector spaces, in particular $\Omega_\st^n=\Omega^n$ for any degree $n$.

It is easy to show that the usual exterior derivative is compatible with
the new $\wedge_\st$-product,
\eq
d(\tau\wedge_\star \tau')=d(\tau)\wedge_\star \tau'+(-1)^{deg(\tau) } \tau\wedge_\star
d\tau'
\en
since the exterior derivative commutes with the  Lie
derivative.

We also have compatibility with the usual undeformed integral (graded cyclicity property):
        \eq
       \int \tau \wedge_\star \tau' =  (-1)^{deg(\tau) deg(\tau')}\int \tau' \wedge_\star \tau\label{cycltt'}
       \en
       with $deg(\tau) + deg(\tau')\!=d=\, $dim$\:\!M$.
       In fact we have, up to boundary terms,
$$       \int \tau \westar \tau' =    \int \tau \wedge \tau'=
(-1)^{deg(\tau) deg(\tau')}\int \tau' \wedge \tau=
(-1)^{deg(\tau) deg(\tau')}\int \tau' \westar \tau
$$
For example at first order in $\theta$,
$$
\int \tau \westar \tau' =    \int \tau \wedge \tau'-{i\over
 2}\theta^{IJ}\int{\cal L}_{X_I}(\tau\wedge {\cal L}_{X_J}\tau')
=
\int \tau \wedge \tau'-{i\over
 2}\theta^{IJ}\int d {i}_{X_I}(\tau\wedge {\cal L}_{X_J}\tau')
$$
where we used the Cartan formula ${\cal L}_{X_I}=di_{X_I}+i_{X_I}d$
and $\tau\wedge {\cal L}_{X_J}\tau'$ being a $d$-form so that its
exterior derivative vanishes.

Finally, provided that the commuting vector fields $\{X_I\}$ 
defining an abelian twist
are all (anti)hermitian, we have compatibility with the undeformed complex conjugation
\eq
       (\tau \wedge_\star \tau')^* =   (-1)^{deg(\tau) deg(\tau')} \tau'^* \wedge_\star \tau^*~.
       \en
Indeed, sending $i$ into $-i$ in the twist (\ref{Mtwist}) amounts to send $\theta^{IJ}$ into
        $-\theta^{IJ} = \theta^{JI}$, i.e. to exchange the
        order of the factors in the $\star$-product.

\section{Noncommutative Yang-Mills actions}

It is straightforward to write a $U(N)$ Yang-Mills theory on
noncommutative space given by Groenewold--Moyal star product,   
\begin{equation}
S_{NCYM}=\,{-1\,\over 2 g^2}^{\,}\int d^4x  \:\label{Action1multiplet}
T^{\!}r(\Fhat_{\mu\nu}\*\Fhat^{\mu\nu})
\end{equation}
where the noncommutative field strength $\Fhat$ is defined by
$$
\widehat F_{\mu\nu}  = \partial_\mu\widehat A_\nu 
- \partial_\nu\widehat A_\mu  -i(\widehat A_\mu\star \widehat A_\nu-\widehat A_\nu\star \widehat A_\mu)
$$
This action
is invariant under the 
noncommutative gauge transformations 
$$\hat\delta \hat A_\mu = \partial_\mu\hat\epsilon
+ i(\hat\epsilon\star\hat A_\mu-\hat A_\mu\star\hat\epsilon)~,
$$
which imply $\hat\delta \widehat F_{\mu\nu}=i(\hat\epsilon\star\widehat F_{\mu\nu}-\widehat
F_{\mu\nu}\star\hat\epsilon)$. Using the differential geometry developed in the previous section this
 action can also be rewritten as 
$$
S_{NCYM}=\,{-1\,\over 2 g^2}^{\,}\int  
T^{\!}r(\Fhat\wedge_\st\ast_{H\!}\Fhat)
$$
where  $A= A_\mu \st dx^\mu$,
$\Fhat=d\Ahat-i\Ahat\wedge_\st\Ahat$ and the Hodge star operator
$\ast_{H\!}$ is the usual commutative one in flat Minkowski metric
(recall that as vector spaces $\Omega_\st^2=\Omega^2$).
The noncommutative gauge transformations now read    
$$\hat\delta \widehat A = d\hat\epsilon
 + i(\hat\epsilon\star\hat A-\hat A\star\hat\epsilon)~.
$$    
and imply $\hat\delta \widehat F=i(\hat\epsilon\star\widehat F-\widehat
F\star\hat\epsilon)$, $\hat\delta (\ast_H\widehat
F)=i\big(\hat\epsilon\star(\ast_H\widehat F)-(\ast_H\widehat F)\star\hat\epsilon\big)$.

In this action the gauge potential and the field strength are valued
in $n\times n$ hermitian matrices, that define the Lie algebra of
$U(N)$. Other representations of $U(N)$ and gauge groups are in general 
problematic. Indeed consider an infinitesimal gauge transformation $\la=\varepsilon^AT^A$, where the
generators $T^A$ belong to some representation 
of a Lie group $G$.
The commutator of two infinitesimal gauge
transformations is
\begin{equation}\label{commgauge}
  [\la\,,~\la']_\st
  \equiv\la\star \la'-\la'\star\la=
{1\over 2}\{\varepsilon^A\,,~\varepsilon'^B\}_\st\,[T^A,T^B]+{1\over
  2}[\varepsilon^A\,,~\varepsilon'^B]_\st\,\{T^A,T^B\}~,
\end{equation}
where $\{U,V\}_\st:=U\star V-V\star U$, $[U,V]_\st:=U\star V-V\star U
$. We see that also the anticommutator $\{T^A,T^B\}$ appears. We thus
have two options:
\\

\noi {\it i)} Consider gauge groups like $U(N)$ or $GL(N)$
in the (anti)fundamental or in the adjoint, since in this case
$\{T^A,T^B\}$ is again in the Lie algebra.\\

\noi {\it ii)} Allow for more general representations of $U(N)$ or $GL(N)$,
or more general Lie algebras (including all simple
Lie algebras)  with representations that 
do not close under the anticommutator. In this case we have to enlarge the Lie
algebra to include also anticommutators besides commutators, i.e., we
have to consider all possible (symmetrized) products $T^AT^B\ldots T^C$ of generators. 
The gauge potential will correspondingly have components $$\hat A=\hat
A^AT^A+\hat A^{AB}T^{AB}+\hat A^{ABC}T^{ABC}+\ldots$$ and therefore
infinite degrees of freedom.  The Seiberg--Witten map discussed in
Section 6 allows to reduce them to the classical degrees of
freedom, and therefore to construct noncommutative Yang--Mills
theories with any gauge group. 

\section{Noncommutative vierbein gravity coupled to fermions}

\subsection{Classical action and symmetries}

 Here we apply the twist procedure to first order gravity in $d=4$
 coupled to fermions and obtain gravity on noncommutative spacetime. 
The usual action of first-order gravity coupled to a spin ${1 \over 2}$
field $\psi$ reads:
 \eq
   S =\epsi_{abcd} \int  R^{ab} \we V^c \we V^d  -
i \bar\psi  \ga^a V^b \we V^c \we V^d \we D\psi 
-i   (D \bar\psi)   \ga^a \we V^b \we V^c \we V^d \psi
  \label{action1comp}
\en
where the vierbein $V^a$ and the spin connection $\omega^{ab}$ are independent one-forms:
\eq
V^a = V^a_\mu dx^\mu,~~~~~\omega^{ab} = \omega^{ab}_\mu dx^\mu
\en
and the two-form (Lorentz) curvature $R^{ab}$ is defined as
\eq
R^{ab} = d\omega^{ab} - \omega^a_{~c} \wedge \omega^{cb}~.
\en
The Dirac conjugate 
is defined as usual: $\psibar = \psi^\dagger \ga_0$. 
 
 This action can be recast
in an index-free notation \cite{Chamseddine}, \cite{AC1}, convenient for generalization to the
noncommutative case:
\eq
 S =  \int Tr \left(i R \we V \we V \ga_5\right)+\psibar   V \we V \we V \ga_5  D\psi +
 D\psibar \we V \we V \we V \ga_5 \psi \label{action1}
\en
\noi where 
 \eq
  R= d\Om - \Om \we
\Om, ~~~~~ D\psi = d\psi - \Om \psi,~~~~~D \psibar =\overline{D\psi}= d \psibar + \psibar  \Om    \label{psipsi}
\en
\noi with
\begin{equation}\label{VaOmab}V  \equiv V^a \ga_a ,~~~~~ \Om  \equiv {1 \over 4} \om^{ab} \ga_{ab} ,~~~~~R \equiv {1\over4} R^{ab} \ga_{ab}
\end{equation}
taking value in Dirac gamma matrices (recalled in Appendix
   \ref{gammamat}). Use of the gamma matrix  identities
$
\ga_{abc} = i \epsi_{abcd} \ga^d \ga_5,$ $
 Tr (\ga_{ab} \ga_c \ga_d \ga_5) = -4 i \epsi_{abcd}
$
 in computing the trace leads back to the usual action
 (\ref{action1comp}).
Reality of the component fields
$V^a$,  $\om^{ab}$, 
is equivalent to the
hermiticity conditions 
\eq
 \ga_0 V \ga_0 = V^\dagger,~~~ -\ga_0 \Omega \ga_0 =
 \Omega^\dagger.~
\label{hermconj}
 \en
The action (\ref{action1}) is
real; for the proof compare it to its complex conjugate, obtained by taking the Hermitian conjugate of the
4-form inside the trace in the integral.

The action
 is invariant under local diffeomorphisms
 (it is the integral of a 4-form on a 4-manifold, hence the
 infinitesimal diffeomorphism ${\cal L}_{X}=di_{X}+i_{X}d$
 reduces to $di_{X}$, a total derivative).
It is also invariant under local Lorentz rotations. These latter read
\eq\label{gaugevariations}
\de_\epsilon V = -[V,\epsilon ] , ~~~\de_\epsilon \Om = d\epsilon - [\Om,\epsilon],~~~~
\de_\epsilon \psi = \epsilon \psi, ~~~\de_\epsilon \psibar = -\psibar \epsilon
\en
\noi with
$
 \epsilon = {1\over 4} \epsi^{ab} \ga_{ab}
$.
The local Lorentz invariance of the index free action follows from
$ \de_\epsilon  R = - [{ R},\epsilon ]$ and $\de_\epsilon D\psi =
\epsilon D\psi$, the cyclicity of the trace $Tr$ and the fact that the
gauge parameter $\epsilon$ commutes with
   $\ga_5$. 

After substituting (\ref{VaOmab}) and $
 \epsilon = {1\over 4} \epsi^{ab} \ga_{ab}
$ into
   (\ref{gaugevariations}), simple gamma algebra yields
   the gauge variations of the component fields:
   \eq
   \delta_\epsilon V^a= \epsi^a_{~b} V^b,~~~\delta_\epsilon \omega^{ab} = d\epsi^{ab} -\omega^a_{~c} \epsi^{cb} + \omega^b_{~c} \epsi^{ca}  \equiv D\epsi^{ab}~.
   \en
   Similarly, the variation of the curvature components is found to be
   \eq
   \delta_\epsilon R^{ab} = \epsi^{a}_{~c} R^{cb} -  \epsi^{b}_{~c} R^{ca}~,
   \en
  while $ \delta_\epsilon (\bar\psi  \ga^a D\psi)=
  \epsi^a_{~b}\bar\psi  \ga^b D\psi$. Thus all quantities in the action (\ref{action1}) transform homogeneously under Lorentz
  local rotations, and since $\epsi_{abcd}$ is an invariant tensor of $SO(1,3)$, the action is
  likewise invariant. Here the proof of invariance looks simple both in the index-free and in the component 
  formulation. Note however that in general the index-free proof is
  much simpler.

\subsection{Noncommutative gauge theory and Lorentz group}\label{4.2Sec}

Before presenting the noncommutative version of the action (\ref{action1}), we discuss the
$\star$-deformation of the Lorentz symmetry variations (\ref{gaugevariations}). These are 
generated by the gamma matrices ${1 \over 4} \gamma_{ab}$, and the gauge parameter is
$\epsilon =  {1\over 4} \epsi^{ab} \gamma_{ab}$. The commutator of two Lorentz transformations
is again a Lorentz transformation, corresponding to the fact that the commutator of two $\gamma_{ab}$ matrices contains only $\gamma_{ab}$ matrices. The situation changes when considering the $\star$-deformation
of this symmetry: as discussed in Section  4, the commutator of two
$\st$-gauge transformations contains
also anticommutators of the generators. The anticommutator of two ${1 \over 4} \gamma_{ab}$ matrices
yields the identity and the $\gamma_5$ matrices, so that the gauge parameter must now include them
in its expansion:
$$\epsilon={1\over
  4}\epsi^{ab}\gamma_{ab}+i\epsi1\!\!1+\tilde\epsi\gamma_5~.$$ 
The extra gauge parameters $\epsi,\tilde\epsi$ can be chosen to
be real (like
$\epsi_{ab}$). Indeed the reality of 
$\epsi_{ab}$, $\epsi,\tilde\epsi$ is equivalent to the
hermiticity condition
\eq\label{hermconde}
 -\ga_0 \epsilon \ga_0 =
 \epsilon^\dagger
\en
and if the gauge parameters $\epsilon$, $\epsilon'$ satisfy this condition then also $ [\epsilon\,\star\!\!\!_,~\epsilon']$
is easily seen to satisfy this
hermiticity condition.

Thus we have centrally extended the Lorentz group to
$$
SO(3,1)\rightarrow SO(3,1)\times U(1)\times R^{+}~,$$ or more precisely,
(since our manifold $M$ has a spin structure and we have a gauge
theory of the spin group $SL(2,C)$)
\[
SL(2,C)\rightarrow GL(2,C)\,.
\]
The Lie algebra generator $i1\!\!1$ is the anti-hermitian generator corresponding to
the $U(1)$ extension, while 
$\gamma_5$ is the hermitian generator  corresponding to the noncompact
$R^+$ extension. 
\sk

Since under noncommutative gauge transformations we have
\eq
\de_\epsilon \Om = d\epsilon - \Om \star \epsilon+ \epsilon
\star \Om
\label{stargauge}
 \en
also the spin connection and the curvature
will  be valued in the
$GL(2,C)$ Lie algebra representation given by all the even gamma matrices,
 \eq
  \Om = {1 \over 4} \om^{ab} \ga_{ab} + i \om 1\!\!1 + \omtilde \ga_5,
  ~~~~~
 R= {1\over 4} R^{ab} \ga_{ab} + i r 1\!\!1 + \rtilde \ga_5~.
\en
Similarly the gauge transformation of the vierbein,
\eq\label{stargauge2}
\de_\epsilon V = -V \star \epsilon + \epsilon \star V,
 \en
closes in the vector space of odd gamma matrices (i.e. the vector
space linearly generated by $\gamma^a,\gamma^a\gamma_5$)
and not in the subspace of just the $\gamma^a$ matrices.
Hence the noncommutative vierbein are  valued in the odd gamma matrices
\eq
V = V^a
\ga_a + \Vtilde^a \ga_a \ga_5  ~.
  \en
Reality of the component fields
$V^a$, $\Vtilde^a$, $\om^{ab}$, $\om$, and $\omtilde$ is equivalent to
the hermiticity conditions 
\eq
 \ga_0 V \ga_0 = V^\dagger,~~~ -\ga_0 \Omega \ga_0 =
 \Omega^\dagger.~
\label{hermconjNC}
 \en
\noi These hermiticity conditions are consistent with the gauge
variations.

  Finally, the infinitesimal gauge transformations of the fields
considered close the Lie algebra of $GL(2,C)$,
   \eq
   [\de_{\epsilon_1},\de_{\epsilon_2}]_\st = -\de_{[\epsilon_1,\epsilon_2]_\st}~.
   \en

\subsection{Noncommutative Gravity action and its symmetries}\label{4.3}

The abelian twist, defining the star products and compatible with usual integration on $M$,
leads to the extension of the Lorentz gauge group  to $GL(2,C)$. It allows to generalize to the noncommutative case the gravity action \eqref{action1}.
The noncommutative action reads
\eq
 S =  \int Tr \left(i {R}\westar \! V \!\westar\! V \ga_5\right) \,+\,
\bar\psi  \star V \!\westar \! V\! \westar \!V\!\westar \!\gamma_5D\psi 
\,+\,  D \bar\psi \westar V \!\westar \! V\! \westar \!V\!\star
\gamma_5\psi
\label{action1NC}
\en
\noi with
 \eq R= d\Om - \Om \westar \Om~,~~~~ D\psi = d\psi -
\Om \star \psi ~,~~~~D \psibar = d \psibar + \psibar  \star \Om    \label{2.20}~.
\en
Reality of this noncommutative action follows by comparing it to its complex conjugate (obtained by taking the Hermitian conjugate of the
4-form inside the trace in the integral).\\

\noi{\bf Gauge invariance}  of the noncommutative action (\ref{action1NC}) under
the $\star$-variations is proved  in  the same way as 
for the commutative case, noting that all the fields in the action transform homogeneously, cf. (\ref{stargauge}),
(\ref{stargauge2}) and
   \eq
\de_\epsilon \psi = \epsilon \star \psi,
~~~\de_\epsilon \psibar = -\psibar \star \epsilon
\label{stargauge3}~,~~
  \de_\epsilon D\psi = \epsilon \star D\psi~,~~
\de_\epsilon D\psibar = -D\psibar \star \epsilon~,~~
  \de_\epsilon R = - R \star \epsilon+ \epsilon \star R~.
   \en
Using that
   $\epsilon$  commutes with $\ga_5$, and  the cyclicity of the
   trace together with the graded cyclicity of the integral, the invariance of (\ref{action1NC}) follows.
\sk
\noi{\bf Diffeomorphisms invariance} 
of the action (up to boundary terms) is proved
straightforwardly by using the Cartan identity $\Lcal_{X}=i_Xd+di_X$ for any
infinitesimal diffeomorfism generated by a vector field $X$ and recalling that the
action is the integral of a $4$-form.  Under these diffeomorphisms the
vector fields $X_I$ transform covariantly:  $\delta_XX_I=
[X,X_I]$.  It is also possible to introduce $\star$-infinitesimal diffeomorphisms \cite{GR2}, see
discussion in \cite{book}, Section 8.2.4. They satisfy a deformed
Leibniz rule and leave invariant the vector fields $X_I$  and
hence the $\star$-product.
\sk
\noi {\bf Charge conjugation invariance.}
Noncommutative charge conjugation reads:
\eq\label{defCconj}
\psi\to\psi^{\;C}\equiv C(\bar\psi)^T=-\gamma_0 C \psi^\ast\!~,~~
V\to V^{\,C}\equiv
C{\,V}^{\;T}C\!~,~~\Omega\to \Omega^{\,C}\equiv 
C{\:\!
\Omega_{}}^{\;T}C\!~
\en
with $\star_\theta\to\star_\theta^C=\star_{-\theta}$ and consequently
$\wedge_{\star_\theta}\to\wedge_{\star_\theta}^{\,C}=\wedge_{\star_{-\theta}}$,
(see Appendix \ref{gammamat} for the properties of the charge conjugation
matrix $C$). 
The action (\ref{action1NC}) is invariant under charge
conjugation. 
\eqa S^C_{bosonic}\!\!&=& i  \int Tr ( {R^C}\wedge_{-\theta} V^C
\wedge_{-\theta} V^C \ga_5 )^T =
- i  \int Tr ( {R^T}\wedge_{-\theta} V^T \wedge_{-\theta} V^T
C\ga_5C^{-1} )^T \nonumber\\
&=& -i \int Tr 
 \left( (V^T \we_{-\theta} V^T \ga_5^T)^T \we_\star R \right)  
= -i \int Tr \left( -(V^T \ga^T_5)^T \we_\star V \we_\star R \right) \nonumber \\ 
& =& i \int Tr ( \ga_5
 V \we_\star  V  \we_\star R) = i \int Tr (R\we_\star  \ga_5
 V \we_\star  V) 
 =   i \int Tr (R\we_\star  
 V \we_\star  V \ga_5) \nonumber \\ 
&=&S_{bosonic}\label{SbosonicC}
\ena
A similar proof holds for the fermionic part of the action : 
$S_{fermionic}^C=S_{fermionic}$. 
\\

\noi{\bf{Noncommutative action and gauge variations for the component fields.}}
Finally, we give the bosonic noncommutative action in terms of the component fields
     $V^a$, $\om^{ab}$, $\Vtilde^a$, $\om$, and
$\omtilde$, and write the gauge variations of these fields.
\eqa
 S_{bosonic} = & &\!\!\!\!\!\!  \int R^{ab} \westar (V^c \westar V^d - \Vtilde^c \westar \Vtilde^d)
 \epsi_{abcd} - 2i~ R^{ab} \westar (V_a \westar  \Vtilde_b - \Vtilde_a \westar
 V_b)\nonumber \\
 & &~~~- 4~r \westar(V^a \westar \Vtilde_a - \Vtilde^a \westar
 V_a)  + 4i~ \rtilde \westar (V^a \westar V_a -\Vtilde^a \westar
 \Vtilde_a)\nonumber \\
\ena

\noi with
 \eqa
  R^{ab}\!\!\!&=&\!\!d \om^{ab} - \unmezzo \om^{a}_{~c} \westar \om^{cb} +
  \unmezzo \om^{b}_{~c} \westar \om^{ca} - i(\om^{ab}
 \westar \om + \om \westar \om^{ab}) - \nonumber \\
  & &~  - {i \over 2}  \epsi^{ab}_{~~cd}( \om^{cd} \westar \omtilde +
   \omtilde \westar \om^{cd}) \nonumber \\
   r\!\!& =&\!\! d\om - {i \over 8} \om^{ab} \westar \om_{ab} -i ( \om \westar
   \om - \omtilde \westar \omtilde )\nonumber \\
     \rtilde\!\! &=&\!\! d\omtilde+ {i \over 16} \epsi_{abcd} \om^{ab} \westar
    \om^{cd}  -i (\om \westar \omtilde + \omtilde
    \westar \om) ~.\nonumber 
     \ena
The noncommutative gauge variations read
\eqa
 & & \de_\epsilon V^a = \unmezzo (\epsi^{a}_{~b} \star V^b + V^b \star
 \epsi^{a}_{~b}) + {i \over 4} \epsi^{a}_{~bcd} (\Vtilde^{b}
 \star \epsi^{cd} -  \epsi^{cd} \star \Vtilde^{b}) \nonumber \\
& &~~~~~~~~~~ + i(\epsi \star V^a - V^a \star \epsi) - \epsitilde \star
\Vtilde^a - \Vtilde^a \star \epsitilde\nonumber \\
 & & \de_\epsilon \Vtilde^a = \unmezzo (\epsi^{a}_{~b} \star \Vtilde^b + \Vtilde^b \star
 \epsi^{a}_{~b}) + {i \over 4} \epsi^{a}_{~bcd} (V^{b}
 \star \epsi^{cd} -  \epsi^{cd} \star V^{b}) \nonumber \\
& &~~~~~~~~~~ + i(\epsi \star \Vtilde^a - \Vtilde^a \star \epsi )-
\epsitilde \star V^a - V^a \star \epsitilde\nonumber \\
 & & \de_\epsilon \om^{ab} = d \epsi^{ab} +\unmezzo (\epsi^a_{~c} \star \om^{cb} -\epsi^b_{~c} \star \om^{ca}
   + \om^{cb} \star  \epsi^a_{~c} - \om^{ca} \star \epsi^b_{~c})
   \nonumber \\
   & & ~~~~~~~~~~ + i (\epsi^{ab} \star \om - \om \star
   \epsi^{ab}) + {i \over 2} \epsi^{ab}_{~~cd} (\epsi^{cd} \star
   \omtilde - \omtilde \star \epsi^{cd})\nonumber \\
   & & ~~~~~~~~~~+ i(\epsi \star \om^{ab} - \om^{ab} \star \epsi)
   + {i \over 2} \epsi^{ab}_{~~cd} (\epsitilde \star
   \om^{cd} - \om^{cd} \star \epsitilde)\nonumber \\
   & & \de_{\epsilon} \om = d \epsi - {i\over 8} (\om^{ab} \star \epsi_{ab} -
   \epsi_{ab} \star \om^{ab}) + i(\epsi \star \om - \om \star \epsi
   - \epsitilde \star \omtilde + \omtilde \star \epsitilde\nonumber )\\
      & & \de_{\epsilon} \omtilde = d\epsitilde +{i \over 16} \epsi_{abcd}
    (\om^{ab} \star \epsi^{cd} - \epsi^{cd} \star \om^{ab}) +
   i( \epsi \star \omtilde - \omtilde \star \epsi + \epsitilde \star
   \om - \om \star \epsitilde)\nonumber
    \ena

\subsection{Classical limit and charge conjugation constraints}
In the classical limit $\theta\rightarrow 0$ the $\star$-product
becomes the usual pointwise product. The noncommutative gauge symmetry
becomes a usual gauge symmetry with gauge group $GL(2,C)$ and the
noncommutative vierbein in the classical limit leads to two
independent vierbeins: $V^a$ and
$\tilde V^a$ transforming both only under the $SL(2,C)$ subgroup of
$GL(2,C)$. As observed in \cite{Chamseddine} this is problematic because we obtain two massless
gravitons and only one local Lorentz symmetry. That is  not enough in
order to kill the unphysical degrees of freedom. Either we concoct a
mechanism such that the second
graviton becomes massive or we further constrain the noncommutative
theory so that in the classical limit the extra vierbein vanishes.

We present two methods of constraining the noncommutative fields. The
first one is based on charge conjugation conditions. The second one
will be presented in Section \ref{NGTAGG}; it
has a wider application and is based on the Seiberg-Witten map. 
\\

The fields in the noncommutative gravity action are in general
$\theta$-dependent as is clear by observing that the $\star$-gauge transformation
of a field is $\theta$ dependent (because of the $\theta$ dependence of
the  $\star$-product). The  vanishing of the $\tilde V^a$ components in the classical limit is
achieved by imposing charge conjugation constraints on the fields \cite{AC1}:
 \begin{equation}
 C V_\theta (x) C = V_{-\theta} (x)^T,~~~C \Omega_\theta (x) C = \Omega_{-\theta} (x)^T,~~~
 C \epsi_\theta (x) C = \epsi_{-\theta} (x)^T \label{ccc}
 \end{equation}
where we have explicitly written the $\theta$-dependence of the fields.
Conditions (\ref{ccc}) are consistent with
$\star$-gauge transformations. For example, the field $ C V_\theta (x)^T C$ can
be shown to transform in the same way as $V_{-\theta} (x)$.

These charge
conjugation constraints imply that the fields and gauge parameter components $V^a,
\om^{ab}, \epsi^{ab}$ are
even in $\theta$, while the components  $\Vtilde^a, \om^{},\omtilde,
\epsi, \epsitilde$
are odd.\\

We then conclude that the
noncommutative gravity action  in \eqref{action1NC}
with fields satisfying the
charge conjugation constraints \eqref{ccc} is real,
diffeomorphisms invariant,  invariant under
$GL(2,C)$ $\star$-gauge transformations and in the classical
limit reduces to the usual gravity action with usual $SL(2,C)$ gauge
invariance. Indeed, only the  fields and gauge parameter components $V^a,
\om^{ab}, \epsi^{ab}$ differ from zero in the classical limit.

As already observed the action is also charge conjugation invariant.
In the presence of the charge conjugation constraints  (\ref{ccc})
the bosonic gravity action is furthermore even in $\theta$.
Indeed  (\ref{ccc}) implies
$
V^{\,C}= 
V_{-\theta},~
\Omega^{\,C}=\Omega_{-\theta},~
R^{\,C}=R_{-\theta}
$.
From the first equality in \eqref{SbosonicC} we see that the bosonic action $S_{bosonic} (\theta) $ is mapped into
$S_{bosonic} (-\theta)$ under charge conjugation, and since it is also
invariant  we  conclude that it is even in $\theta$.

\section{Seiberg--Witten map}\label{SWl}
\def\epsi{\epsilon}
We first study  the Seiberg-Witten map between commutative and noncommutative  gauge
 theories with noncommutativity given by the  Groenewold--Moyal product.
\\

In a gauge theory physical quantities are
gauge invariant: they do not depend on the gauge potential but on
the gauge equivalence class of the potential given. 
The Seiberg-–Witten map relates the noncommutative gauge fields to the
commutative ones by requiring the noncommutative fields to have the
same gauge equivalence classes as the commutative ones
\cite{SW}. 
Explicitly, the noncommutative gauge
potential $\SWA=A_\mu{\rm d}x^\mu$  and the noncommutative gauge
parameters $\epsihat$ depend on  the ordinary $A$ and $\epsi$ so to satisfy:
 \eq
 \Ahat (A + \de_\epsi A) = \Ahat (A) + \dehat_\epsihat \Ahat (A) \label{SWcondition}
 \en
 with 
  \eqa 
   & &
  \de_\epsi A_\mu = \part_\mu \epsi - i A_\mu 
      \epsi+ i \epsi A_\mu, \\
      & &
  \dehat_\epsihat \Ahat_\mu = \part_\mu \epsihat - i \Ahat_\mu \star     
      \epsihat + i \epsihat \star \Ahat_\mu.\label{3.44h}
      \ena
This equation  can be solved order by order in powers of the
noncommutativity parameter $\theta$
yielding
$\Ahat$ and $\epsihat$ as  power series in $\theta$:  
 \eqa 
   & & \Ahat (A) = A + A^1 (A)  + A^2 (A) + \cdots + A^n (A)+ \cdots  \\
    & & \epsihat (\epsi, A)  =  \epsi + \epsi^1 (\epsi, A)+ \epsi^2 (\epsi, A)+ \cdots + 
     \epsi^n (\epsi, A)+ \cdots 
     \ena    
 \noi  where $A^n (A)$ and $\epsi^n (\epsi, A)$  are of order $n$ in $\theta$. Note that  $\epsihat$
depends on the ordinary $\epsi$ and also on $A$.
For example, up to first order it is readily checked that
   \eqa
& & \Ahat_\kappa = A_\kappa+A_\kappa^1(A)+O(\theta^2)= A_\kappa-
 \frac{\theta^{\mu\nu}}{4} \{ A_{\mu}, \part_{\nu } A_\kappa +
 F_{\nu\kappa} \}   +O(\theta^2)  \label{one} \\
& &  \epsihat =\epsi+\epsi^1+  O(\theta^2)=\epsi -
 \frac{\theta^{\mu\nu}}{4} \{  A_{\mu}, \part_{\nu} \epsi\} +O(\theta^2)\label{oneepsi}
 \ena
  with $F_{\mu\nu} := \part_\mu A_\nu -  \part_\nu A_\mu - i A_\mu  A_\nu
    + i A_\nu A_\mu$ and where  $   \{U,V\}= U V + VU$ is the
    anticommutator of two operators.
    
The Seiberg--Witten condition (\ref{SWcondition}) holds for any value of the
noncommutativity parameter $\theta$.
If we consider it at $\theta'$ and at $\theta$ we easily obtain 
that gauge equivalence classes of the $\theta'$-noncommutative theory
have to correspond to gauge equivalent classes of the
$\theta$-noncommutative theory, i.e., we generalize
(\ref{SWcondition}) to
\begin{equation}\label{SWconditionp}
\SWAp(\SWA+\SWV_{\SWE} \SWA)=\SWAp(\SWA) -
\SWVp_{\!\!\;\SWEp}\SWAp(\SWA)~,
\end{equation}
where we denoted by $\star'$, $\SWAp$, $\SWEp$, $\SWVp_{\!\!\:\SWEp}$
the star product, the gauge potential, the gauge parameter and the gauge variation: 
$\SWVp_{\!\!\;\SWEp}\SWAp_\kappa=\partial_\kappa\SWEp-i\SWAp_\kappa\star'\SWEp+i\SWEp\star'\SWAp_\kappa$
at noncommutativity parameter $\theta'$.
By considering $\theta$ and $\theta'$ infinitesimally close, so that
$\theta' = \theta + \delta \theta$ and $\SWAp = \SWA + \delta
\theta^{\mu\nu} \frac{\partial \SWA}{\partial \theta^{\mu\nu}}$ (we
consider $\frac{\partial}{\partial \theta^{\mu\nu}}$
independent from $\frac{\partial}{\partial \theta^{\nu\mu}}$ and hence
sum over all $\mu$,$\nu$ indices) a rather straightfoward
computation, generalizing that for \eqref{one} and \eqref{oneepsi},
shows that if $\Ahat$ and
$\epsihat$  solve the differential equations 
   \eqa
 & & { \part \over \part \theta^{\mu\nu}} \Ahat_\kappa = -
 \frac{1}{8} \Big(\{ \Ahat_{\mu}, \part_{\nu } \Ahat_\kappa +
 \Fhat_{\nu\kappa} \}_\star    -   \{ \Ahat_{\nu}, \part_{\mu } \Ahat_\kappa +
 \Fhat_{\mu\kappa} \}_\star \Big)   \label{one1} \\
 & &  { \part \over \part \theta^{\mu\nu}} \epsihat =  -
 \frac{1}{8} \Big(\{  \Ahat_{\mu}, \part_{\nu} \epsihat \}_\star  -  \{  \Ahat_{\nu}, \part_{\mu} \epsihat \}_\star\label{two}\Big)
 \ena
  where $\{U,V\}_\star= U\star V + V \star U$ and  
  \begin{equation}
\Fhat_{\mu\nu} := \part_\mu \Ahat_\nu -  \part_\nu \Ahat_\mu - i \Ahat_\mu \star \Ahat_\nu
    + i \Ahat_\nu \star \Ahat_\mu ~,\label{Fhatmunu} \\
    \end{equation}
then $\SWAp(\SWA)$ and $\SWEp(\SWE,\SWA)$
satisfy also the Seiberg--Witten condition 
(\ref{SWconditionp}) for arbitrary  values of $\theta'$ and
$\theta$. In
particular, therefore, they solve the Seiberg--Witten condition (\ref{SWcondition}).

   The differential equations (\ref{one1}) and (\ref{two}) admit solutions in terms of formal power series in $\theta$. These are given recursively by
    \eqa
     & &
     A^{n+1}_\mu = -{1\over 4(n+1)} \theta^{\rho\sigma} \{\Ahat_\rho, \part_\sigma \Ahat_\mu   
     + \Fhat_{\sigma\mu} \}^n_\star\,, \label{An+1}\\
      & & \epsi^{n+1}=  -{1 \over 4(n+1)} \theta^{\rho\sigma} \{\Ahat_\rho, \part_\sigma \epsihat \}^n_\star\,,\label{rtwo}
       \ena
   \noi where  $ \{ \fhat, \ghat \}^n_\star $ is the $n$-th  order term in $ \{ \fhat, \ghat \}_\star $, so that for example
     \begin{equation}
     \{\Ahat_\rho, \part_\sigma \epsihat \}^n_\star \equiv \sum_{r+s+t=n} (A^r_\rho \star^s \part_\sigma \epsi^t +
      \part_\sigma \epsi^t \star^s A^r_\rho )\,.
       \end{equation}
    \noi Here $\star^s$ indicates the $s$-th order term in the star
    product expansion  \cite{Ulker}. There is a  simple proof of
    \eqref{An+1}, \eqref{rtwo} \cite{AC1}:
   multiplying the differential equations by $\theta^{\mu\nu}$ and analysing them
   order by order yields
   \eqa & &\theta^{\mu\nu}  { \part \over \part \theta^{\mu\nu}} A_\rho^{n+1} = (n+1) A_\rho^{n+1}= - \frac{1}{4}  \theta^{\mu\nu}  \{ \Ahat_{\mu}, \part_{\nu } \Ahat_\rho + \Fhat_{\nu\rho} \}_\star^n~,~~\nonumber\\
   & &\theta^{\mu\nu}  { \part \over \part \theta^{\mu\nu}} \epsi^{n+1}=  (n+1) \epsi^{n+1}= - \frac{1}{4} \theta^{\mu\nu}  \{  \Ahat_{\mu}, \part_{\nu} \epsihat \}_\star^n\nonumber
  \ena
\noindent since $A_\rho^{n+1}$ and $\epsi^{n+1}$ are homogeneous functions of $\theta$ of order 
$n+1$. 
\sk
Similar considerations hold
for matter fields $\phi$  transforming
in the fundamental or in the adjoint representation of the gauge group.
The Seiberg--Witten condition reads, cf. \cite{Jurco:2001rq},
\begin{equation}\label{SWconditionmatter}
\phihat(A+\delta_\epsi 
A, \phi+\delta_\epsi \phi) =\phihat(A,\phi)+\hat\delta_\epsihat\phihat(A,\phi)\,,
\end{equation}
or more generally,
\begin{equation}\label{SWconditionmatterp}
\SWPp(\SWA+\delta_{\SWE} \SWA, \SWP+\delta_{\SWE} \SWP) =\SWPp(\SWA,
\SWP)+\SWVp_{\SWEp}\SWPp(\SWA,\SWP)\,,
\end{equation}
and it is satisfied if the matter fields solve the differential equation
\begin{equation}\label{matteradj}
  \begin{split}
 &\delta \theta^{\mu\nu} \frac{\partial \SWP}{\partial \theta^{\mu\nu}}
 =-\frac{1}{4} \delta\theta^{\mu\nu}\SWA_\mu \star
 (\partial_\nu \SWP + D_\nu \SWP )\; ~~~~~~~ {\rm
   fundamental ~rep.,~ i.e., ~}\dehat_\epsihat \phihat = i \epsihat \star \phihat\,, ~~~~
\\ &\delta \theta^{\mu\nu} \frac{\partial \SWPA}{\partial \theta^{\mu\nu}}
 =-\frac{1}{4} \delta\theta^{\mu\nu}\bigl\{\SWA_\mu \,, (\partial_\nu
 \SWPA + D_\nu \SWPA )\bigr\}_\star\;  ~~ {\rm
   adjoint ~rep.,~ i.e.,~ }\dehat_\epsihat \Psihat = i \epsihat \star \Psihat - i \Psihat \star \epsihat\,. 
\end{split}
\end{equation}
The explicit solutions order by order in $\theta$ are
\eqa     & & \phi^{n+1}= -{1 \over 4(n+1)} \theta^{\mu\nu} \big(\Ahat_\mu \star (\part_\nu \phihat
     + D_\nu \phihat) \big)^{\!\:\!n} ~~~~~~~~~~{\rm (fundamental)}\label{phirec} \\
 & & \Psi^{n+1}= -{1\over 4(n+1)} \theta^{\mu\nu} \{\Ahat_\mu, \part_\nu \Psihat
     + D_\nu \Psihat \}^n_\star ~~~~~~~~~~~~~~{\rm (adjoint)}
\label{phiadrec}
\ena
where
$$
D_\nu \phihat = \part_\nu \phihat -i  \Ahat_\nu \star \phihat ~~,~~~~ 
D_\nu \Psihat = \part_\nu \Psihat -i [ \Ahat_\nu , \Psihat]_\star 
$$
are the covariant derivative in the fundamental 
and  in the adjoint, with $[S,T]_\star:=S\star T-T\star S$.
\\

The Seiberg--Witten differential equations \eqref{one1}, \eqref{two},
\eqref{matteradj} are not the most general solutions to the gauge
equivalence condition \eqref{SWcondition}. For example, from the
differential equation for the gauge potential it is easy to see that the field
strength $\hat F_{\mu\nu}$ satisfies the differential equation
$$
 \delta \theta^{\rho\sigma} \frac{\partial \Fhat_{\mu\nu}}{\partial \theta^{\rho\sigma}}  = -{1 \over 4} \theta^{\rho\sigma} \big( \{\Ahat_\rho, \part_\sigma \Fhat_{\mu\nu}   
     + D_\sigma \Fhat_{\mu\nu} \}_\star - 2\{\Fhat_{\mu\rho},\Fhat_{\nu\sigma} \}_\star \big) 
$$
which has an extra addend with respect to the differential equation \eqref{matteradj} for fields transforming in the adjoint. The most general Seiberg--Witten differential
equations are presented in Appendix \ref{sec:ambig}.
The freedom in the  Seiberg--Witten differential
equations may be useful for their integration, see e.g. Appendix \ref{sec:ambig}.

\subsection{Geometric Seiberg-Witten map}
The  Seiberg--Witten map considered for Groenewold--Moyal
noncommutativity can be generalized to the case of an abelian twist
\begin{equation}
{\cal F}= e^{-\frac{i}{2}\theta^{IJ}X_I \otimes X_J} \label{Abeliantwist}
\end{equation}
where $\{X_I\}$ is a set of mutually commuting vector fields globally
defined on a manifold $M$ and $\theta^{IJ}$ is a constant
antisymmetric matrix. The corresponding $\star$-product is obtained
composing  the usual pointwise
multiplication $\mu (f \otimes g)= fg$ with the inverse twist
${\cal F}^{-1}= e^{\frac{i}{2}\theta^{IJ}X_I \otimes X_J}$,
\begin{eqnarray}
f\star g &=& \mu  ({\cal F}^{-1} (f\otimes g)) ~.
\end{eqnarray}
This $\star$-product is in general
position dependent because the vector fields $X_I$ are in general
$x$-dependent. Associativity of the $\star$-product is guaranteed by
mutual commutativity of the vector fields $X_I$.  In the special case that $M=\mathbb{R}^d$ and
$X_I={\partial \over \partial x^I}$, $I=1,...,d$ we recover 
the Groenewold--Moyal $\star$-product (\ref{GMW}). 
 
The use of vector fields $\{X_I\}$ on a manifold
$M$ suggests a coordinate independent approach to the Seiberg--Witten
map. The resulting NC gauge potential $\hat A$ is then
a $1$-form that depends on $A$,
on the mutually  commuting vector fields $X_I$ and on the deformation matrix
$\theta=(\theta^{IJ})_{I,J=1,...,d}$.
The coordinate independent expression of the Seiberg--Witten
differential eq.s (\ref{one}),(\ref{two}) reads
  \eqa
 & & { \part \over \part \theta^{IJ}} \Ahat= - \unquarto \{ i_{X_{[I}} \Ahat, \Lcal_{X_{J]}} \Ahat+  i_{X_{J]}} \Fhat \}_\star\label{onegeom} \\
 & &  { \part \over \part \theta^{IJ}} \epsihat =  - \unquarto \{  i_{X_{[I}} \Ahat, \Lcal_{X_{J]}} \epsihat \}_\star\label{twogeom}
 \ena
\noi where $\Fhat \equiv d\Ahat - i \Ahat \westar \Ahat$ is a two-form,
$i_{X_I}$ and $ \Lcal_{X_I}$  are respectively the contraction and the Lie derivative along the mutually
commuting vector fields $X_I$. When the abelian twist reduces to the Groenewold--Moyal case of  the preceding Section, the curvature becomes
 $\Fhat = \unmezzo  \Fhat_{\mu\nu} dx^\mu \we dx^\nu$, where the $ \Fhat_{\mu\nu}$ components are given in
(\ref{Fhatmunu}). Note that in the Groenewold--Moyal case $dx^\mu \westar dx^\nu
=dx^\mu \we dx^\nu$ since $\Lcal_{X_I} dx^\mu = d \Lcal_{X_I}x^\mu=0$.
Proceeding as in the Groenewold--Moyal case the recursive solutions are given by:
 \eqa
     & &
     A^{n+1} = -{1 \over 4(n+1)} \theta^{IJ} \{i_{X_I} \Ahat, \Lcal_{X_J} \Ahat
     + i_{X_J}  \Fhat \}^n_\star \label{Asol}\\
      & & \epsi^{n+1}=  -{1 \over 4(n+1)} \theta^{IJ} \{ i_{X_I} \Ahat, \Lcal_{X_J}  \epsihat \}^n_\star\label{epsisol}
       \ena
Similarly one proves  the generalization of eq.s (\ref{phirec})-(\ref{phiadrec}):
\eqa
      \phi^{n+1}&\!\!=\!\!& -{1 \over 4(n+1)} \theta^{IJ} \left( i_{X_I} \Ahat  \star (2 \Lcal_{X_J} \phihat
      -i (i_{X_J}    \Ahat) \star \phihat ) \right)^n, ~~\dehat_\epsihat \phihat = i \epsihat \star \phihat \label{phirecgen} \\
 \Psi^{n+1}&\!\!=\!\!& -{1 \over 4(n+1)} \theta^{IJ} \{  i_{X_I} \Ahat \,, \,2\Lcal_{X_J} \hat\psi
      - i (i_{X_J}\Ahat) \star \SWPA + i \SWPA \star (i_{X_J}\Ahat)\}_\star^n, ~~\dehat_\epsihat \SWPA = i \epsihat \star \SWPA- i\SWPA\star\epsihat\nonumber\\
\label{phiadrecgen}
\ena

In this subsection we have constructed the geometric Seiberg--Witten map for noncommutative gauge
theories with gauge group $U(N)$ or $GL(N)$ (or products thereof) and
with $\star$-product given by a general abelian 
twist \eqref{Abeliantwist}.   For abelian gauge groups the
Seiberg--Witten map can be constructed for any
$\star$-product associated with an arbitrary Poisson tensor.
The map is obtained in \cite{JSW} using Kontsevich formality theorem \cite{Kont}.
The study of its global geometric aspects shows that the
Seiberg--Witten map quantizes line bundles with connections on a
Poisson manifold to quantum (noncommutative) line bundles  with  noncommutative
connections \cite{Jurco:2001kp}
The Seiberg--Witten map for nonabelian gauge groups and with arbitray
Poisson tensors is in general an open problem, we refer to  \cite{Jurco:2001my}
for interesting insights.
The global geometric aspects of the Seiberg–Witten map are well understood for nonabelian $U(n)$-gauge
fields on noncommutative tori:  the Seiberg–Witten map defined in
\eqref{CDD}, \eqref{CDD'}, with $\gamma=-3$, $\rho=i$, quantizes vector bundles on tori
with connections to vector bundles on noncommutative tori with
noncommutative connections and the results are nonformal in the sense
that they do not rely on power series expansion in the noncommutativity
parameter $\theta$. 

We also mention that the Seiberg-Witten map for Chern--Simons gauge
theories can be studied to all orders in the noncommutativity
parameter $\theta$ \cite{Vilar}, \cite{Aschieri:2014xka}.

\section{Noncommutative gauge theories with any gauge group}\label{NGTAGG}

Up to now we have considered noncommutative gauge theories with gauge
group $U(N)$ or $GL(N)$ in the fundamental or adjoint, and more
generally representations of gauge groups such that the
generators $T^A$ of the Lie algebra close also in the usual matrix
product. This is needed for the closure of infinitesimal gauge transformations, cf. \eqref{commgauge}.
If on the other hand we consider an arbitrary gauge group $G$ equation
\eqref{commgauge} shows that infinitesimal gauge transformations do not
close in the Lie algebra
Lie$(G)$, but in the universal enveloping algebra ${\mathcal{U}}($Lie$(G))$. This latter
is the product of all generators $T^A$ modulo the relations $T^AT^B-T^BT^A=[T^A,T^B]$.
When considering an arbitrary gauge group $G$ the noncommutative gauge potential is
therefore universal enveloping algebra valued
$$
\Ahat=\Ahat^AT^A+\Ahat^{AB}T^AT^B+\Ahat^{ABC}T^AT^BT^C+\ldots~
$$
and hence with infinitely many (symmetric) components $\Ahat^A$,
$\Ahat^{AB}$, $\Ahat^{ABC}$, .... . The Seiberg--Witten map is  well
defined also in this case. 
It then constrains these infinite components 
to depend on the commutative ones $A^A=A^A_\mu dx^\mu$ in $A=A^AT^A$.
Similarly, the gauge parameters and the matter fields  
depend on the commutative gauge parameters and matter fields, besides
the commutative gauge potential.
\\

This Seiberg--Witten map approach to noncommutative gauge
theories is  called universal enveloping algebra valued approach \cite{Jurco:2001rq}. It has been used to propose
noncommutative standard and grand unified particle physics
models \cite{Calmet:2001na, Aschieri:2002mc} having the same  degrees of freedom as in the commutative models. Renormalizability and scattering amplitudes using the Seiberg--Witten map have been studied e.g. in
\cite{Buric, Schupp:2008fs, Martin:2009vg, Horvat:2011qg}, either considering
a power series expansion in $\theta$
or a  $\theta$-exact approach (i.e. to all arders in $\theta$)
\cite{Schupp:2008fs, Martin12}, where the power series is instead in the gauge
coupling constant.
For recent literature on scattering amplitudes of noncommutative
particle models using the
Seiberg--Witten map see \cite{Trampetic:2021awu} and references therein.
\\

We next study the Seiberg--Witten map for the gravity action \eqref{action1NC}.

\subsection{Expansion of gravity fields to first  order in $\theta$}

Up to first order in $\theta$ the solution
to the Seiberg-Witten conditions for the gravity fields reads:
\eqa
& &
\hat\Omega=\Omega-\frac{1}{4}\theta^{IJ}\{\Omega_I\label{omsol}
,  {\cal L}_J \Omega+  R_{J} \} +\mathcal{O}(\theta^2)\\
& & \epsihat= \epsi-{1 \over 4} \theta^{IJ} \{\Om_I, {\cal L}_J \epsi \} +\mathcal{O}(\theta^2)\\
 & & \hat\psi= \psi-{1 \over 4} \theta^{IJ} \Om_I  ( {\cal L}_J + L_J) \psi +\mathcal{O}(\theta^2)\label{solSW}\\
& & \hat R = R-{1 \over 4} \theta^{IJ} \left( \{\Om_I, ( {\cal L}_J + L_J)  R  \} 
 - [R_{I},R_{J} ] \right) +\mathcal{O}(\theta^2) \label{Rsol}
 \ena
where
$\Om_I$, $R_I$ are defined as the contraction along the tangent
vector $X_I$ of
the exterior forms $\Om$, $R$, i.e., $\Om_I\equiv i_I\Om$,
$R_I \equiv i_I R$, ($i_I$ being the contraction along $X_I$).
We have also introduced the Lie derivative ${\cal L}_I$ along the vector
field $X_I$, and the covariant Lie derivative $L_I$ along the
vector field $X_I$. $L_I$ acts on $R$ and $\psi$ as 
$L_I R={\cal L}_I R- i\Om_I \star
R+ i R \star\Om_I$ and
$L_I  \psi= {\cal L}_I \psi - \Om_I \psi $.
In fact the covariant Lie derivative $L_I$ has the Cartan form:
 \eq
  L_I = i_I D + D i_I~\nonumber
    \en
where $D$ is the covariant derivative.
We refer to \cite{AC2} for higher order in $\theta$ expressions.

\subsection{Expansion of noncommutative gravity action at first order in
  $\theta$}
The Seiberg--Witten map allows to expand the noncommutative action in
terms of the commutative fields. Noncommutative gravity is therefore
seen as commutative gravity with further interaction terms due to
noncommutativity of space time.

The Seiberg--Witten solutions  \eqref{omsol}-\eqref{Rsol} are {\it not} $SO(1,3) $-gauge covariant, due to
the presence of the ``naked" connection
$\Om$ and the non-covariant Lie derivative ${\cal L}_I=i_I{\dd}+{\dd}
i_I$. 
However, when inserted in the NC action the resulting action is gauge invariant
order by order in $\theta$. Indeed usual gauge variations induce the
$\star$-gauge variations under which the noncommutative action is
invariant.
Therefore the NC action, re-expressed in terms of ordinary
 fields via the SW map, is invariant under usual gauge
 transformations.
Moreover the action, once re-expressed in terms of ordinary fields remains geometric,
and hence invariant under diffeomorphisms. This is the case because the noncommutative
action and the SW map are geometric: indeed only
coordinate independent operations like the contraction $i_I$ and the
Lie derivatives ${\cal L}_I$ and $L_I$ appear in the Seiberg--Witten map.

We replace the noncommutative fields appearing in the action
with their expansions (\ref{omsol})-(\ref{Rsol}) in commutative fields, 
and integrating by parts we obtain
the following gravity action coupled to spinors
\eqa\label{action11}
S &\!\!=\!\!&\!\!  \int Tr \left(i R V V \ga_5\right)+\psibar   V^3\ga_5  D\psi +
 D\psibar V^3 \ga_5 \psi \\ 
&
&+\frac{i}{4}\theta^{IJ}\Big(\bar\psi\{V^3,R_{IJ}\}\gamma_5D\psi+D\bar\psi\{V^3,R_{IJ}\}\gamma_5\psi\Big)\nonumber\\[.5em]
& &+\frac{i}{2}\theta^{IJ}\Big(2L_I\bar\psi
R_JV^3\gamma_5\psi-2\bar\psi V^3R_I\gamma_5L_J\psi - L_I\bar\psi V^3\gamma_5L_JD\psi-L_ID\bar\psi\,
V^3\gamma_5 L_J\psi \nonumber\\
&&+\bar\psi (\{L_I\!VL_J\!V,V\}+L_I\!V\, V L_J\!V)\gamma_5D\psi +D\bar\psi  (\{L_I\!VL_J\!V,V\}+L_I\!V\, V L_J\!V)\gamma_5\psi 
\Big)
+O(\theta^2)\nonumber
\ena
where we have omitted writing the wedge product, and
$V^3=V\wedge V\wedge V$.
The expression of the gravity action, up to second order in $\theta$,
in terms of the commutative fields has been given in \cite{ACD}, after
a propaedeutical study of the Seiberg--Witten map for $\star$-products of fields.

\section{Conclusions}

We have constructed a gravity action in noncommutative spacetime --an
expected feature of quantum spacetime-- and
shown its equivalence to the usual gravity action (in the first order
formalism) on commutative spacetime with extra interaction terms.
These are obtained from spacetime noncommutativity using the
Seiberg--Witten map between commutative and noncommutative gauge
theories.  This extended gravity action 
is  invariant under local Lorentz transformations because it is
expressed solely in terms of gauge covariant operators $L_I, i_I,
D$, curvature $R$, vierbein $V$, spinor fields $\psi$, invariant vector
fields   $\{X_I\}$ and noncommutativity parameter $\theta$.
It is diffeomorphic invariant and charge conjugation invariant.
This noncommutative  gravity action can also be coupled to
noncommutative scalar and gauge fields \cite{ACdyn, ACGG}. Choosing
an appropriate kinetic term the vector
fields   $\{X_I\}$ can become dynamical, the idea being that
both spacetime curvature and noncommutativity should depend on  matter
distribution. It would be interesting to study cosmological models as solutions of
these extended gravity actions.
 
\appendix

\section{Ambiguities in the Seiberg--Witten map}\label{sec:ambig}

The solution to the Seiberg--Witten conditions (\ref{SWcondition}), (\ref{SWconditionmatter}) is not unique. For example
      if $\hat A_\mu$ is a solution, any  noncommutative gauge
      transformation of $\hat A_\mu$ gives another solution. Another
      source of ambiguities is that of field redefinitions of the
      gauge potential (e.g., if $\hat A_\mu$ is a solution then so is
      $\hat A_\mu+\theta^{\rho\sigma}\theta^{\lambda\eta}\hat F_{\rho
        \lambda}\star D_\sigma \hat F_{\eta\mu}$).
We generalize the 
Seiberg--Witten equations  (\ref{one}),
(\ref{two}) and (\ref{matteradj}) allowing for three extra terms
$\hat D_{\mu\nu\rho}(\SWA),  \hat E_{\mu\nu}(\SWA, \SWE)$,
$\hat C_{\mu\nu}(\SWA, \SWP)$ and $\hat C_{\mu\nu}(\SWA, \SWPA)$ that
are a priori arbitrary functions of their arguments and derivatives thereof, that are (formal) power series in $\theta$ and 
that are antisymmetric in the $\mu,\nu$
indices. We consider the equations
\begin{eqnarray}
  \label{extendedSWA}
  \!\!\!\!\!\delta^\theta \SWA_\kappa\,=\, \delta \theta^{\mu\nu} \frac{\partial
  \SWA_\kappa}{\partial \theta^{\mu\nu}}&\!\!
                                          =&\!\!-\frac{1}{4}\delta\theta^{\mu\nu}\Bigl(\{\SWA_\mu,\partial_\nu
                                             \SWA_\kappa +
                                             \SWF_{\nu\kappa}\}_\star
                                             +
                                             \hat D_{\mu\nu\kappa}(\SWA)\Bigr)\,,~~~~~~\\
\label{extendedSWE}
   \!\!\!\!\! \delta^\theta\SWE\,=\,\delta \theta^{\mu\nu}
  \frac{\partial\SWE}{\partial \theta^{\mu\nu}} &\!\!=&\!\!-\frac{1}{4}
                                                      \delta\theta^{\mu\nu}\Bigl(\{\partial_\mu
                                                        \SWE ,
                                                        \SWA_\nu\}_\star
                                                        +
                                                       \hat  E_{\mu\nu}(\SWA,\SWE)\Bigr)\,,\\
\label{extendedSWP}
    \!\!\!\!\!  \delta^\theta\SWP\,=\,\delta \theta^{\mu\nu} \frac{\partial
  \SWP}{\partial \theta^{\mu\nu}} &\!\!=&\!\!-\frac{1}{4}
                                          \delta\theta^{\mu\nu}\Bigl(\SWA_\mu
                                          \star \partial_\nu \SWP +
                                          \SWA_\mu \star D_\nu \SWP +
                                          \hat
                                          C_{\mu\nu}(\SWA,\SWP)\Bigr)\,,\\
   \!\!\!\!\!   \delta^\theta\SWPA\,=\,\delta \theta^{\mu\nu} \frac{\partial
  \SWPA}{\partial \theta^{\mu\nu}} &\!\!=&\!\!-\frac{1}{4}
                                          \delta\theta^{\mu\nu}\Bigl(\{\SWA_\mu
                                          , \partial_\nu \SWPA +
                                           D_\nu \SWPA\}_\star
                                           +
                                          \hat
                                           C_{\mu\nu}(\SWA, \SWPA)\Bigr)
\label{extendedSWPA}
\end{eqnarray}
and observe that $\hat E_{\mu\nu}$ must be linear in $\SWE$ since all terms in
(\ref{extendedSWE}) but $\hat E_{\mu\nu}$ are linear in $\SWE$,  similarly
$\hat C_{\mu\nu}$ must be linear in $\SWP$ because of the linearity in
$\SWP$ of all other terms in (\ref{extendedSWP}), and similarly for
$\hat C_{\mu\nu}(\SWA, \SWPA)$ in \eqref{extendedSWPA}. Imposing the Seiberg--Witten conditions  (\ref{SWconditionp}),
(\ref{SWconditionmatterp}) we obtain the conditions
\begin{equation}
  \begin{split}
\hat D_{\mu\nu\kappa}(\SWA+\SWV_{\SWE} \SWA)-\hat
D_{\mu\nu\kappa}(\SWA) - i[\SWE,\hat D_{\mu\nu\kappa}(\SWA)]_{\star}
&=-  D_\kappa \hat E_{\mu\nu}(\SWA,\SWE)\;,
\\
\label{constrC}
\hat C_{\mu\nu}(\SWA+\SWV_{\SWE}\SWA,\SWP+\SWV_{\SWE}\SWP) -
 \hat C_{\mu\nu}(\SWA,\SWP) - i\SWE\star \hat C_{\mu\nu}(\SWA,\SWP) &=
 -i\hat E_{\mu\nu}(\SWA,\SWE)\star \SWP\;,\\
\hat C_{\mu\nu}(\SWA+\SWV_{\SWE}\SWA,\SWPA+\SWV_{\SWE}\SWPA) -
 \hat C_{\mu\nu}(\SWA,\SWPA)  -i[\SWE,\hat C_{\mu\nu}(\SWA,\SWPA)]_\star &=
  -i[\hat E_{\mu\nu}(\SWA,\SWE),\SWPA]_\star\;.
\end{split}
\end{equation}
In particular we notice that any $\hat D_{\mu\nu\kappa}$ and $\hat C_{\mu\nu}$ covariant under gauge transformations solve  (\ref{constrC}) with
$\hat E_{\mu\nu}=0$. 

In summary, as discussed in \cite{Aschieri:2018vgu}, the most general solution
$\SWA(A)$, $\SWE(A, \varepsilon)$, $\SWP(A,\phi)$, $\SWPA(A,\Psi)$ of the Seiberg--Witten
conditions (\ref{SWcondition}), (\ref{SWconditionmatter}) is given by the differential equations 
(\ref{extendedSWA})-(\ref{extendedSWPA}) where $\hat D,\hat E, \hat C$
are constrained by  (\ref{constrC}).
Further constraints on the $\hat D,\hat E, \hat C$ terms are obtained
by requiring that the Seiberg--Witten map respects hermiticity and charge conjugation in the sense that
the hermiticity and charge conjugation properties of the commutative
fields imply those of the noncommutative fields \cite{Aschieri:2002mc, AC2}.
If we ask  $\hat D,\hat E, \hat C$ to have no explicit dependence on
$\theta$ and to be covariant under constant $GL(d,\mathbb{R})$
coordinate transformations (the star product  $f\star g$ is itself
invariant under constant $GL(d,\mathbb{R})$ coordinate transformations: $x^\mu\to
M^\mu{}_\rho x^\rho$, $\theta^{\mu\nu}\to M^\mu{}_\rho M^\nu{}_\sigma
\theta^{\rho\sigma}  $) we recover the results in
\cite{Asakawa:1999cu} and in \cite{Suo:2001ih}: 
\begin{align}\label{DEsemplice}
&\hat D_{\mu\nu\kappa}=\alpha D_\kappa \SWF_{\mu\nu}+\beta
D_\kappa[\SWA_\mu,\SWA_\nu]_\star~~,~~~~\hat E_{\mu\nu}=2\beta [\partial_\mu\SWE,\SWA_\nu]_\star\;,\\
&\label{Cfund}\hat C_{\mu\nu} = -2i\beta[\SWA_\mu,\SWA_\nu]_\star \star \SWP + \gamma
  \SWF_{\mu\nu}\star \SWP 
~~~~~~~~~~~~~~~~~~~~~~~~~~~~~~~~~~{\rm (fundamental)}\;
\\
&\label{Cadj}\hat C_{\mu\nu} = -2i\beta[[\SWA_\mu,\SWA_\nu]_\star, \SWPA]_\star + \gamma'
  \SWF_{\mu\nu}\star \SWPA +\tilde\gamma \SWPA \star \SWF_{\mu\nu}
~~~~~~~~~~\:~~~~~~ {\rm (adjoint)}\;
\end{align}
with $\alpha,\beta,\gamma, \gamma'$ and $\tilde\gamma$ arbitrary constants.

An interesting Seiberg--Witten differential equation is obtained
considering \eqref{extendedSWA}-\eqref{extendedSWPA} with $\hat D_{\mu\nu\kappa}=0$, $  \hat
E_{\mu\nu}=0$ and
\begin{align}
  &{\hat C_{\mu\nu}(\hat A,
   \hat\phi)= \gamma
   \SWF_{\mu\nu}\star \SWP + \rho D_\mu D_\nu \SWP
   ~~~~\mbox{ { for $\mu<\nu$,}
                                  and }~ \hat C_{\nu\mu}:=-\hat C_{\mu\nu}\label{CDD}}
   \\[.2em]                               &{\hat C_{\mu\nu}(\hat A, \hat\Psi) = \gamma
   [\SWF_{\mu\nu}, \SWPA]_\star + \rho D_\mu D_\nu \SWPA
   ~\mbox{ { for $\mu<\nu$,}
   and }~ \hat C_{\nu\mu}:=-\hat C_{\mu\nu}\label{CDD'}}
\end{align}
Here $\rho$ is a constant and
 $GL(d,\mathbb{R})$ covariance is broken because   $D_\mu D_\nu \SWPA$
 is not antisymmetric in the $\mu, \nu$ indices, i.e., $\hat C_{\nu\mu}$ does not contain
also the term $\rho D_\nu D_\mu\phi$.
This choice, with
$\gamma=-3$ and $\rho=i$, allows to solve the Seiberg--Witten map 
on  noncommutative tori (obtained from the Groenewold--Moyal
noncommutative plane) to all orders in $\theta$ for topologically nontrivial
$U(N)$-gauge potentials with constant field strengths \cite{Aschieri:2018vgu}.

\section{Gamma matrices in $D=4$}\label{gammamat}

We summarize in this Appendix our gamma matrix conventions in $D=4$.
\begin{flalign}
&  \eta_{ab} =(1,-1,-1,-1),~~~\{\ga_a,\ga_b\}=2 \eta_{ab},~~~[\ga_a,\ga_b]=2 \ga_{ab}, \\
&  \ga_5 \equiv i \ga_0\ga_1\ga_2\ga_3,~~~\ga_5 \ga_5 = 1,~~~\varepsilon_{0123} = - \varepsilon^{0123}=1, \\
&  \ga_a^\dagger = \ga_0 \ga_a \ga_0, ~~~\ga_5^\dagger = \ga_5 \\
&  \ga_a^T = - C \ga_a C^{-1},~~~\ga_5^T = C \ga_5 C^{-1}, ~~~C^2 =-1,~~~C^\dagger=C^T =-C
\end{flalign}

\noi{\bf{Useful identities}}
\begin{flalign}
 & \ga_a\ga_b= \ga_{ab}+\eta_{ab}\\
 &  \ga_{ab} \ga_5 = {i \over 2} \varepsilon_{abcd} \ga^{cd}\\
 & \ga_{ab} \ga_c=\eta_{bc} \ga_a - \eta_{ac} \ga_b -i \varepsilon_{abcd}\ga_5 \ga^d\\
 & \ga_c \ga_{ab} = \eta_{ac} \ga_b - \eta_{bc} \ga_a -i \varepsilon_{abcd}\ga_5 \ga^d\\
 & \ga_a\ga_b\ga_c= \eta_{ab}\ga_c + \eta_{bc} \ga_a - \eta_{ac} \ga_b -i \varepsilon_{abcd}\ga_5 \ga^d\\
 & \ga^{ab} \ga_{cd} = -i \varepsilon^{ab}_{~~cd}\ga_5 - 4 \de^{[a}_{[c} \ga^{b]}_{~~d]} - 2 \de^{ab}_{cd}\\
&  Tr(\ga_a \ga^{bc} \ga_d)= 8~ \de^{bc}_{ad} \\
&  Tr(\ga_5 \ga_a \ga_{bc} \ga_d) = -4 i\,\varepsilon_{abcd} 
 \end{flalign}
\sk
\noi where
$\delta^{ab}_{cd} \equiv
\frac{1}{2}(\delta^a_c\delta^b_d-\delta^b_c\delta^a_d)$skyp
and indices antisymmetrization in square brackets has total weight $1$. 
\\

\noi {\large \bf Acknowledgements}\\
The authors acknowledge partial support from INFN, CSN4, Iniziativa
Specifica GSS.  This research has a financial 
support from Universit\`a del Piemonte Orientale. P.A. is
affiliated to INdAM-GNFM.

\end{document}